\documentclass[pre,amsmath,amssymb,twocolumn,superscriptaddress,showpacs]{revtex4-1}

\usepackage{amsmath,amssymb}
\usepackage[usenames]{color}
\usepackage{amssymb}
\usepackage{grffile}
\usepackage[pdftex]{graphicx}
\usepackage{amsmath, amstext, amssymb, amsfonts, amsxtra}
\usepackage{textcomp}
\usepackage{xspace}
\usepackage{bbm}

 \usepackage[utf8]{inputenc}
 \usepackage[colorlinks,
	linkcolor=blue,
	citecolor=blue,
	urlcolor=blue]{hyperref}

\def \tr{{\mbox{tr~}}}

\def \ell{{d}}

\newcommand{\Hop}{\hat{H}}

\newcommand{\nop}{\hat{n}}

\newcommand{\aop}{\hat{a}^{\phantom{\dagger}}}
\newcommand{\adop}{\hat{a}^{\dagger}}
\newcommand{\Dop}{\mathcal{D}}

\newcommand{\lan}{\langle}
\newcommand{\ran}{\rangle}

\newcommand{\im}{{\rm i}} 
\newcommand{\rhop}{\hat{\rho}} 
\newcommand{\rhops}{\hat{\rho}_{\mbox{\tiny ps}}} 
\newcommand{\Hops}{\hat{H}_{\sigma}}
\newcommand{\Hopp}{\hat{H}_{\rm a}}
\newcommand{\Hopi}{\hat{H}_{\sigma{\rm a}}}

\newcommand{\sop}{\hat{\sigma}}

\newcommand{\cursp}{\mathcal{J}^{\sigma}_T}
\newcommand{\curpp}{\mathcal{J}^{\rm n}_{T}} 
\newcommand{\Lind}{\mathcal{L}}  
\newcommand{\Lf}{\mathbb{L}} 

\newcommand{\Pow}{\mathcal{P}}

\newcommand{\bra}[1]{\langle #1|}
\newcommand{\ket}[1]{|#1\rangle}
 
\newcommand{\refe}[1]{Eq.~(\ref{EQ:#1})}

\newcommand{\reff}[1]{Fig.~\ref{FIG:#1}}

\newcommand{\sutd}{Singapore University of Technology and Design, 8 Somapah Road, 487372 Singapore} 
\newcommand{\como}{Center for Nonlinear and Complex Systems, Dipartimento di Scienza e Alta Tecnologia,
Universit\`a degli Studi dell'Insubria, via Valleggio 11, 22100 Como, Italy} 
\newcommand{\infn}{Istituto Nazionale di Fisica Nucleare, Sezione di Milano, via Celoria 16, 20133 Milano, Italy}    
\newcommand{\brazil}{International Institute of Physics, Federal University of Rio Grande do Norte, Natal, Brazil}   
\newcommand{\NEST}{NEST, Istituto Nanoscienze-CNR, I-56126 Pisa, Italy}

\begin{document}

\title{Minimal motor for powering particle motion from spin imbalance}       
\author{Ulf Bissbort} 
\affiliation{\sutd}
\author{Colin Teo} 
\affiliation{\sutd}  
\author{Chu Guo}
\affiliation{\sutd} 
\author{Giulio Casati} 
\affiliation{\como} 
\affiliation{\brazil}  
\author{Giuliano Benenti}  
\affiliation{\como} 
\affiliation{\infn}   
\affiliation{\NEST}   
\author{Dario Poletti}
\affiliation{\sutd}

\begin{abstract}
We introduce a minimalistic quantum motor for coupled energy and particle transport. The system is composed of two spins, each coupled to a different bath and to a particle which can move on a ring consisting of three sites. We show that the energy flowing from the baths to the system can be partially converted to perform work against an external driving, even in the presence of moderate dissipation. We also analytically demonstrate the necessity of coupling between the spins. We suggest an experimental realization of our model using trapped ions or quantum dots. 
\end{abstract}       

\pacs{05.70.Ln,05.60.Gg,03.65.Yz}      

\maketitle

\section{Introduction} \label{sec:intro}  
Systems out of equilibrium can be considered a resource which can be exploited in order to extract useful work. A prime example are thermodynamic engines in which part of the heat flowing from a hot to a cold bath can be converted into work. This is a particularly fascinating challenge for future nanotechnologies where quantum mechanics is necessary for an accurate description (for reviews see, e.g. Refs.~\cite{GiazottoPekola2006, Shakouri2011, Dubi2011, Hanggi2011, Sothmann2014, Benenti2016, MuhonenPekola2012, Seifert2012, Kosloff2013, Gelbwaser2015, Vinjanampathy2015, Benenti2016b}).
Aspects of this quest include the study of the role of coherence and entanglement, quantum measurements, minimum temperature achievable in small quantum chillers, quantum statistics and quantum fluctuations as well as feedback effects \cite{AlickiJenkins2016,GoupilLecoeur2016} and engineered
non-equilibrium distributions for the baths \cite{AlickiNJP2015}.  

Important examples of energy conversion devices are thermoelectric motors. 
A key point in understanding the functioning of such devices is the emergence of different types of coupled energy flows, such as phononic and electronic transport. It is thus important to study systems with coupled transport on a fundamental level, casting particular focus on the energy conversion performance and efficiency.

A useful approach to uncover fundamental principles for improving the performance of energy conversion, is to study minimal models in which different aspects can be more systematically isolated and analyzed. So far, minimal models for heat engines have been used ~\cite{ScovilSchulzDuBois1959, HenrichMichel2007, Youssef2009, Linden2010, Skrzypczyk2011, Gelbwaser-Klimovsky2013, TeoPoletti2016, RouletScarani2016} to study steady-state heat transfer and conversion from thermal reservoirs at different temperatures.  

On the other hand, for the purposes of energy conversion in nanodevices, such as nanoscale thermoelectric devices, it is necessary to consider coupled flows.
In Sec.~\ref{sec:model} we introduce a minimal motor of coupled flows, namely of coupled spin and particle transport. The motor is composed of two coupled spins, each connected to its own bath that dissipatively tends to `pump' the spin into a statistical mixture of $z$-eigenstates with adjustable polarization. To this minimal chain with spin imbalance imposed by the baths, we connect the smallest possible circuit consisting of a particle hopping between three sites, as depicted in \reff{fig1}. In Sec.~\ref{sec:working} we study the conversion of the energy flow between the baths to generate a particle current, and hence power, against a force generated by the external driving. 
We demonstrate that tuning the coupling between the spins strongly affects the current and the power generated and we show the robustness of the motor by studying its performance in presence of dephasing. We also find analytical necessary conditions for the energy conversion to be possible and discuss the (non-)separability of the density operator.  
In Sec.~\ref{sec:conclusions} we draw our conclusions, comment on the baths used and discuss possible experimental realizations with ultracold ions or quantum dots.

\begin{figure}
\includegraphics[width=\columnwidth]{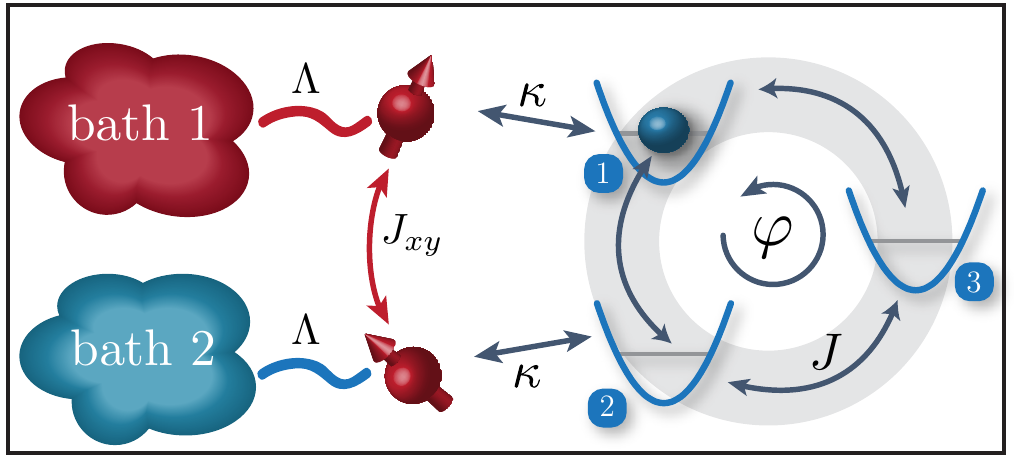}
\caption{(color online) Schematic depiction of the minimal motor: Two spins, each coupled to a different bath and to a particle which can move between three different sites on a ring, generating a current. The particle is exposed to a time-periodic gauge field $\phi(t)$, mimicking a homogeneous force around this minimal circuit.}
\label{FIG:fig1} 
\end{figure}

\section{Model} \label{sec:model}
We consider a system (see \reff{fig1}) composed of two $1/2$-spins and one particle which moves in a minimal circuit consisting of three sites. 
The evolution of the system is described by a master equation in Lindblad form, acting on the density operator $\rhop\,$:
\begin{align}      
\frac{d\rhop}{dt}=\Lind(\rhop,t)=-\frac{\im}{\hbar}[\Hop(t),\rhop] + \Dop(\rhop).  \label{EQ:masterequation}
\end{align}       
The Hamiltonian $\Hop$ is given by  
\begin{align} 
&\Hop =\Hops+\Hopp+\Hopi, \label{eq:Ham}\\      
&\Hops = -J_{xy} \left(\sop_{x,1}\sop_{x,2}+\sop_{y,1}\sop_{y,2}\right) + \sum_l h_{z,l} \, \sop_{z,l}, &\nonumber\\ 
&\Hopp = -J \sum_{l} e^{-\im \phi(t)}\adop_l\aop_{l+1}+\mbox{H.c.}, \nonumber\\ 
&\Hopi = \sum_{l=1,2}\kappa_l \sop_{z,l}\nop_l,  \nonumber    
\end{align}   
where $\Hops$ and $\Hopp$ are the spin and particle Hamiltonians respectively, while $\Hopi$ is the coupling between the spins and the particle. Periodic boundary conditions are implicitly understood for the particle. $\sop_{i,l}$ is the Pauli spin operator for $i=x,y,z$,  acting on the $l-$th spin, $\aop_l$ ($\adop_l$) annihilates (creates) a particle at site $l$ and $\nop_l=\adop_l\aop_l$ counts the number of particles at site $l$; $J_{xy}$ is the $XY$-coupling strength between the spins; $2 h_{z,l}$ sets the Zeeman energy splitting; $J$ is the hopping strength (henceforth we work in units of $J=\hbar=1$); and $\kappa_l$ is the spin-particle coupling strength. 
The particle hopping also includes a time-dependent phase $\phi(t)=\phi(t+T)$ which models the effect of an external periodic driving, analogous to the effect of a time-dependent magnetic field on a charged particle. 

The system undergoes dissipative dynamics, with the dissipator $\Dop=\Dop_{\lambda,1}+\Dop_{\lambda,2}$ in \refe{masterequation} describing the coupling of the spins to the baths. We choose the spin dissipator to be of form
\begin{align} 
\Dop_{\lambda,l}(\rhop)=&  \left[\lambda^+_l \left(2\sop_l^+ \rhop \sop_l^- -\sop_l^-\sop_l^+\rhop - \rhop\sop_l^-\sop_l^+ \right)\right.\nonumber\\ 
&\left. +\lambda^-_l \left(2\sop_l^- \rhop \sop_l^+ -\sop_l^+\sop_l^-\rhop - \rhop\sop_l^+\sop_l^- \right)\right] \label{eq:bd}  
\end{align}  
for each spin, as used in Refs.~\cite{BenentiRossini2009, SchuabLandi2016} for example. The $\lambda_l^{\pm}$ are the spin raising or lowering rates for the spins and are associated to the spin raising and lowering operators $\sop_l^{\pm}=(\sop_{x,l}\pm \im \sop_{y,l})/2$ respectively. To parametrize the system more conveniently, we introduce the total bath coupling strength $\Lambda_l=\lambda_l^+ + \lambda_l^-$, the relative pumping rate into the upper state $p_l=\lambda_l^+/\Lambda_l$  for each spin and the average relative pumping rate $\bar{p}=(p_1+p_2)/2$, which determines the total polarization $\sum_l\langle\sigma_{z,l}\rangle$. Notice that if a spin is exclusively coupled to one bath ($J_{xy}=\kappa_1=\kappa_2=0$), it is driven towards a mixed state, described by the (reduced) density operator
\begin{align} 
\rhop_{\mbox{\tiny spin, $l$}}=p_l \ket{\uparrow}_l\bra{\uparrow}+(1-p_l) \ket{\downarrow}_l\bra{\downarrow}, 
\label{EQ:spinDM_decoupled}
\end{align} 
where $\ket{\uparrow}_l$ and $\ket{\downarrow}_l$ are eigenstates of $\sop_{z,l}$. 

For the sake of simplicity,
we focus on the symmetric scenario $\Lambda_1=\Lambda_2=\Lambda$, $\kappa_1=\kappa_2=\kappa$, and $h_{z,1}=h_{z,2}=h_{z}$. 
A spin current is generated by an imbalance in the baths $\Delta p = p_2-p_1$ and this, as we will show below, can be used to generate power.

\section{Working of the quantum motor} \label{sec:working}     
The system sets a particle into motion against an external driving by converting part of the energy flowing from one bath to the other. 
We consider a periodic driving of the sawtooth form $\phi(t)=\mbox{mod}(vt,2\pi)$, such that $\phi(t+T)=\phi(t)$ with a period  $T=2\pi/v$ \cite{sinedriving}. 
The Lindbladian has the same periodicity, $\Lind(t)=\Lind(t+T)$, and for $t \to \infty$ the system relaxes to a periodic steady state $\rhops(t)$, which we numerically found to be unique for $\kappa\ne 0$. 
To determine the steady state, we numerically propagate a basis of the density operator space over one period. This yields the Floquet Lindbladian operator $\Lf_{t_0}=\mathcal{T}e^{\int_{t_0}^{t_0+\tau} \Lind(t) dt}$ (with $\mathcal{T}$ being the time ordering operator), which is not unitary, but possesses an eigenvalue with value 1. The associated eigenvector is the steady state $\rhops(t_0+nT)$, where $n$ is an integer number.    

An important quantity for the analysis of our model is the particle current averaged over one period,
\begin{align}
{\curpp}=\frac{1}{T}\int_{t_0}^{t_0+T}\tr\!\left[\rhops(t) \; \hat \j_l \right] dt,      
\end{align}   
which is independent of the site $l$ and of the initial time $t_0$. The density operator $\rhops(t)$ is obtained by evolving $\rhops(t_0)$ for time $t-t_0$ using \refe{masterequation} 
\cite{HartmannHanggi2016, totalrho}, while $\hat \j_{l}= \im   
[e^{\im \phi(t)}\adop_{l+1}\aop_{l} - e^{-\im \phi(t)}\adop_{l}\aop_{l+1}]$ is the particle current operator associated with the continuity equation of the local particle number $\nop_l$. Analogously, the spin current averaged over one period is given by 
\begin{align}
{\cursp}=\frac{1}{T}\int_{t_0}^{t_0+T}\tr\!\left[\rhops(t) \; \hat \j^{\sigma} \right] dt,      
\end{align}   
where $\hat \j^{\sigma}_{l}= \im  (2J_{xy})  
\left(\sop_{y,1}\sop_{x,2} - \sop_{y,2}\sop_{x,1}\right)$ is the spin current operator associated with the continuity equation of the local magnetization $\sop_{z,l}$.  

Differentiating the system's internal energy $E(t)=\langle H(t)\rangle = \tr\!\left\{\hat{H}(t)\rhops(t)\right\}$ with respect to time, we obtain an energy balance equation,
$d E/dt =  \dot{Q} - \dot{W}$, where $\dot{W}$ and $\dot{Q}$ are defined as $\dot{W}(t)=-\tr\!\left\{ \frac{\partial \hat{H}(t)}{\partial t}\rhops(t)\right\}$ and $\dot{Q}(t)=\tr\!\left\{\hat{H}(t)  \frac{\partial \rhops(t)}{\partial t}\right\}$ \cite{alicki79, kosloff84, kosloff94}.
Note that $\dot{Q}(t)=0$ for unitary evolution while the power $\Pow(t)=\dot{W}(t)$ can be non-zero only if the Hamiltonian parameters change in time.  
From the definition of the dissipative contribution to the master equation \refe{masterequation}, $\dot{Q}(t)$ is composed of two terms $\dot{Q}(t)=\sum_{i=1,2}\dot{Q}_{\lambda,i}(t)$, where we defined
\begin{equation}
\dot{Q}_{\lambda,i}(t)=
\tr\!\left\{ \hat{H}(t) \, \mathcal{D}_{\lambda,i}\left[\rhops(t)\right]\right\}. 
\end{equation}
It is thus possible to identify $\dot{Q}_{\lambda,i}$ as the energy exchanged with the $i-th$ bath (i.e., a heat current) and $\Pow$ with the work done per unit time (power) by the system against the periodic driving. A positive $\dot{Q}$ indicates that the system absorbs a net amount of energy from the baths. Similarly, the power $\Pow$ is positive when the system performs work, i.e., energy leaves the system. 

We use the notation $A_T=\frac 1 T\int_{t_0}^{t_0+T} A(t)\, dt$ to indicate a time average for any quantity $A(t)$ associated with $\rhops(t)$. If the system is in the periodic steady state, $\rhops(t)=\rhops(t+T)$, then one can relate the motor's average power to the average rate of net energy flowing into the system 
\begin{equation}
\Pow_T=\dot Q_T.
\end{equation}
For our specific choice of driving $\phi(t)$, we can directly relate the power to the particle current
\begin{align} 
\Pow_T=3 v \curpp=\frac{6\pi\curpp}{T}. \label{eq:power}
\end{align}
 
Hence, as the current changes direction, the power and the net energy exchanged with the baths change sign. 

To quantify the efficacy of our motor in transforming heat into work, we define, for positive average powers, the efficiency
\begin{align} 
\eta=\frac{T \, \Pow_T}{Q_{\rm abs}}
\end{align}
 as the ratio of the work performed and the heat $Q_{\rm abs}$ absorbed from the baths over one period. To obtain a fair quantification of the amount of heat absorbed by the system, we account for all contributions instantaneously flowing into the system from either bath 
$Q_{\rm abs}=\sum_{i=1,2}\int_{t_0}^{t_0+T} \dot{Q}_{\lambda,i}(t) \: {\rm \Theta}\!\!\left[\dot{Q}_{\lambda,i}(t)\right]  dt$, where  ${\rm \Theta}(x)$ denotes the Heaviside function \cite{direction}. 
In general it is necessary to have a detailed knowledge of the density operator in order to measure $Q_{\rm abs}$.

\begin{figure}
\includegraphics[width=\columnwidth]{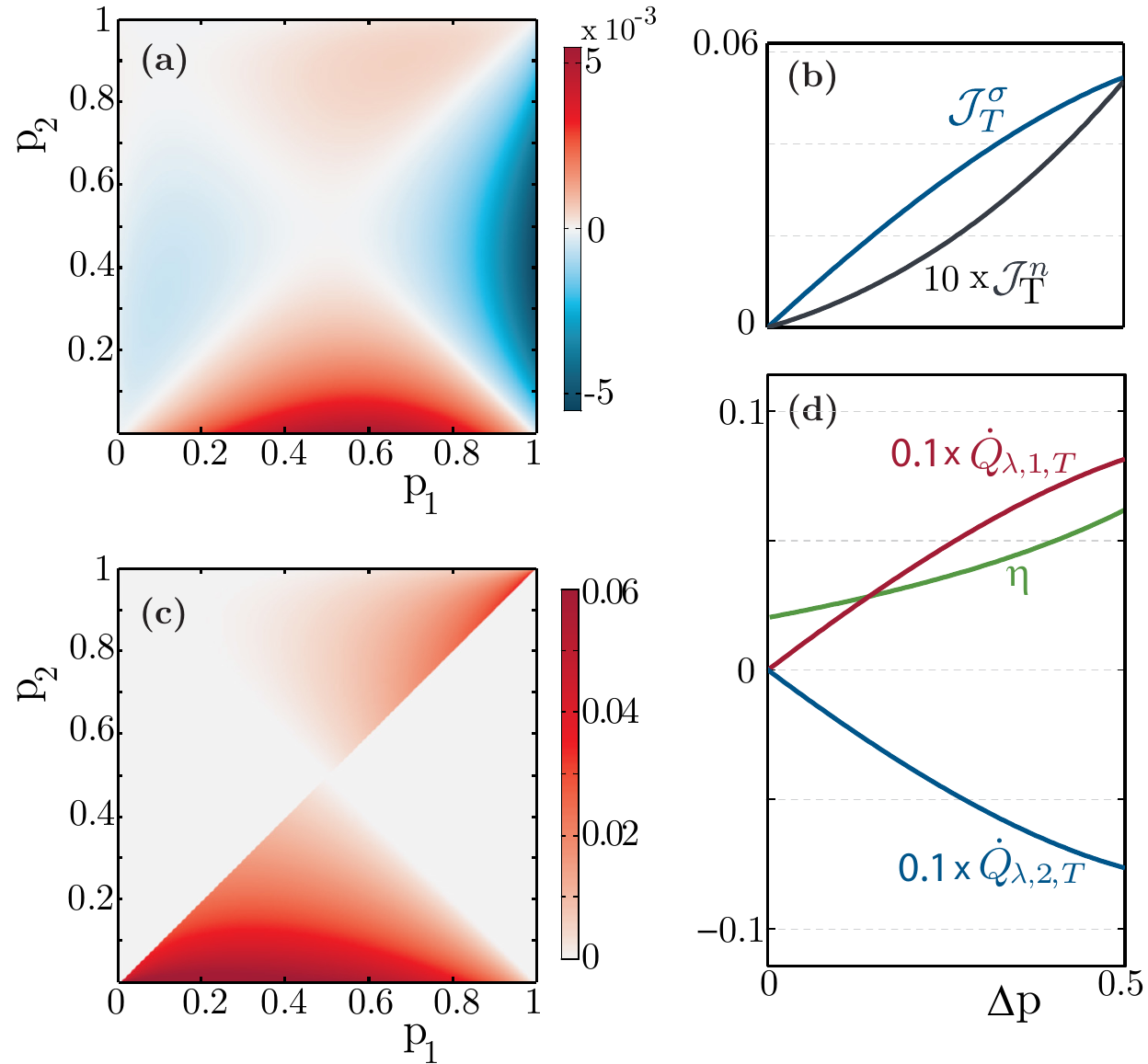}
\caption{(a) Particle current $\curpp$ and (c) efficiency for different values of the bath parameters $p_1$ and $p_2$ [gray areas in (c) correspond to regions with negative current/power in (a), where the system cannot be used for power production]. (b) Particle $\curpp$ and spin currents $\cursp$ versus the baths imbalance $\Delta p$. (d) Efficiency $\eta$ and average energy exchange rates $\dot{Q}_{\lambda,l,T}$ versus the baths imbalance $\Delta p$. Parameters: $\kappa=-2.5$, $J_{xy}=0.3$, $h_z=15$,  $\Lambda=0.2$, $T=2$ and for (b) and (d) $\overline p=0.25$. } \label{FIG:fig2}    
\end{figure}   

In \reff{fig2} we study the particle current and the efficiency of the system as a function of the bath pumping rates $p_l$ and their difference $\Delta p$. In \reff{fig2}(a) we depict the particle current for different values of $p_1$ and $p_2$. Generally, the current vanishes for $p_1=p_2$ and features a linear dependence on $\Delta p$ for small $\Delta p$, as shown in \reff{fig2}(b). The spin current has an analogous behavior close to $\Delta p=0$. 
The qualitative behavior of the efficiency differs from that of the particle current [see Figs.~\ref{FIG:fig2}(c) and \ref{FIG:fig2}(d)]. In particular, for $\Delta p\rightarrow 0$, the efficiency approaches a non-zero value.

\begin{figure}
\includegraphics[width=\columnwidth]{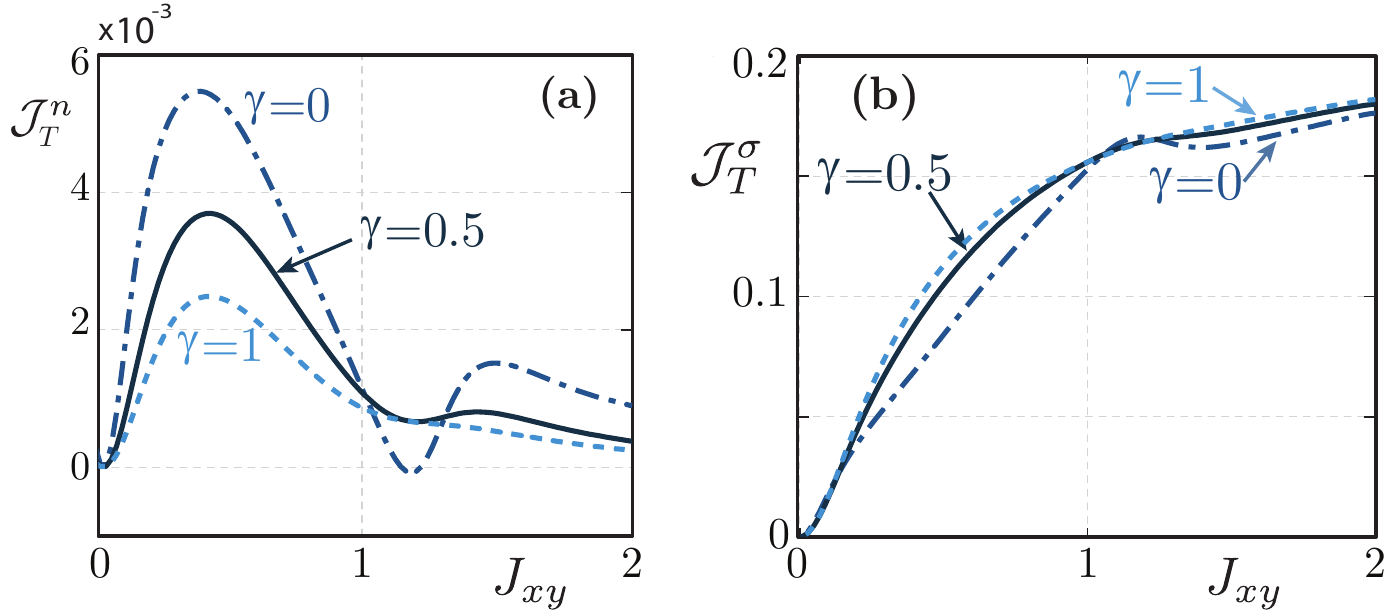}
\caption{(a) Particle current $\curpp$ and (b) spin current $\cursp$ vs. spin coupling $J_{xy}$ (dot-dashed line $\gamma=0$; continuous line, $\gamma=0.5$; and dashed line, $\gamma=1$). 
Common parameters:  $\kappa=-2.5$, $h_z=15$, $\Lambda=0.2$, $\Delta p=0.5$, $\overline{p}=0.25$, and period $T=2$.
}
\label{FIG:fig3}    
\end{figure}

\subsection{Robustness to dephasing} \label{sec:robust}    
We now demonstrate the robustness of the current to a local dephasing dissipative term
\begin{align} 
\Dop_{\gamma}= &  \gamma \left(2\nop_3 \rhop \nop_3 -\nop_3^2\rhop - \rhop\nop_3^2 \right),  \label{eq:dephasing}
\end{align}
where $\gamma$ is the dephasing rate \cite{GardinerZoller}. Then the dissipator $\Dop$ of Eq.~(\ref{EQ:masterequation}) becomes $\Dop=\Dop_{\lambda,1}+\Dop_{\lambda,2}+\Dop_{\gamma}$. Moreover, $\dot{Q}=\dot{Q}_{\lambda,1}+\dot{Q}_{\lambda,2}+\dot{Q}_{\gamma}$, where 
\begin{align}
\dot{Q}_{\gamma}(t)&=\tr\!\left\{ \hat{H}(t) \, \mathcal{D}_{\gamma}\left[\rhops(t)\right]\right\}. 
\end{align}   
The local dephasing term $\Dop_{\gamma}$, which acts only on the third site, mimics a local resistor and tends to suppress the particle delocalization (coherence) $\lan \adop_{l+1}\aop_{l} \ran$ and thus the current. Qualitatively similar behavior persists in presence of global dephasing. In a cold atom setup these dissipations can be engineered using non-far detuned lasers inducing spontaneous emissions \cite{GerbierCastin2010, PichlerZoller2010}.

Figure~\ref{FIG:fig3}(a) shows the particle current as a function of the spins coupling parameter $J_{xy}$. The particle current $\curpp$ is zero for $J_{xy}=0$ and shows a non-monotonic behavior which is robust against dissipation. The spin current $\cursp$, shown in \reff{fig3}(b), is also robust against dephasing. It is important to notice that while $J_{xy}$ needs to be non-zero to have particle current (see below), the fraction of energy converted into power decreases as $J_{xy}$ increases. 
There is hence an optimal value of the power generated versus the spin coupling $J_{xy}$. 

\subsection{Energy transfers}   

Here we elaborate on the various energy flows through the systems via the baths and the time-dependent driving. In \reff{fig4} these contributions are shown as functions of the particle dephasing rate $\gamma$ for typical parameters, as used in the manuscript. The top (red) and bottom curves (blue) show the net energy (averaged over one period $T$) exchanged with the first and second magnetization bath, respectively. $\dot{Q}_{\lambda,1,T}>0$ indicates that energy enters the system from the first bath, while $\dot{Q}_{\lambda,2,T}<0$ shows that energy leaves the system into the second bath. Similarly, the dissipative particle dephasing term $\mathcal D_\gamma$ has an associated net energy flow $\dot{Q}_{\gamma,T}$, which has a non-monotonous behavior in $\gamma$, even changing sign and is about an order of magnitude smaller than the work performed by the system (note the scaling factor 10 in \reff{fig4} for clarity). The average power $\mathcal P_T$ performed by the system (green curve) decreases with $\gamma$.  

It is interesting to note that the magnitude of both energy transfers with the baths $\dot{Q}_{\lambda,1,T}, \dot{Q}_{\lambda,2,T}$ increase in magnitude with $\gamma$; however, their sum (second line from the top) decreases. The net power output $\mathcal{P}_T$ (third line from the top) and the efficiency are thus lowered by the dephasing.  

\begin{figure}[ht]
\includegraphics[width=0.8\columnwidth]{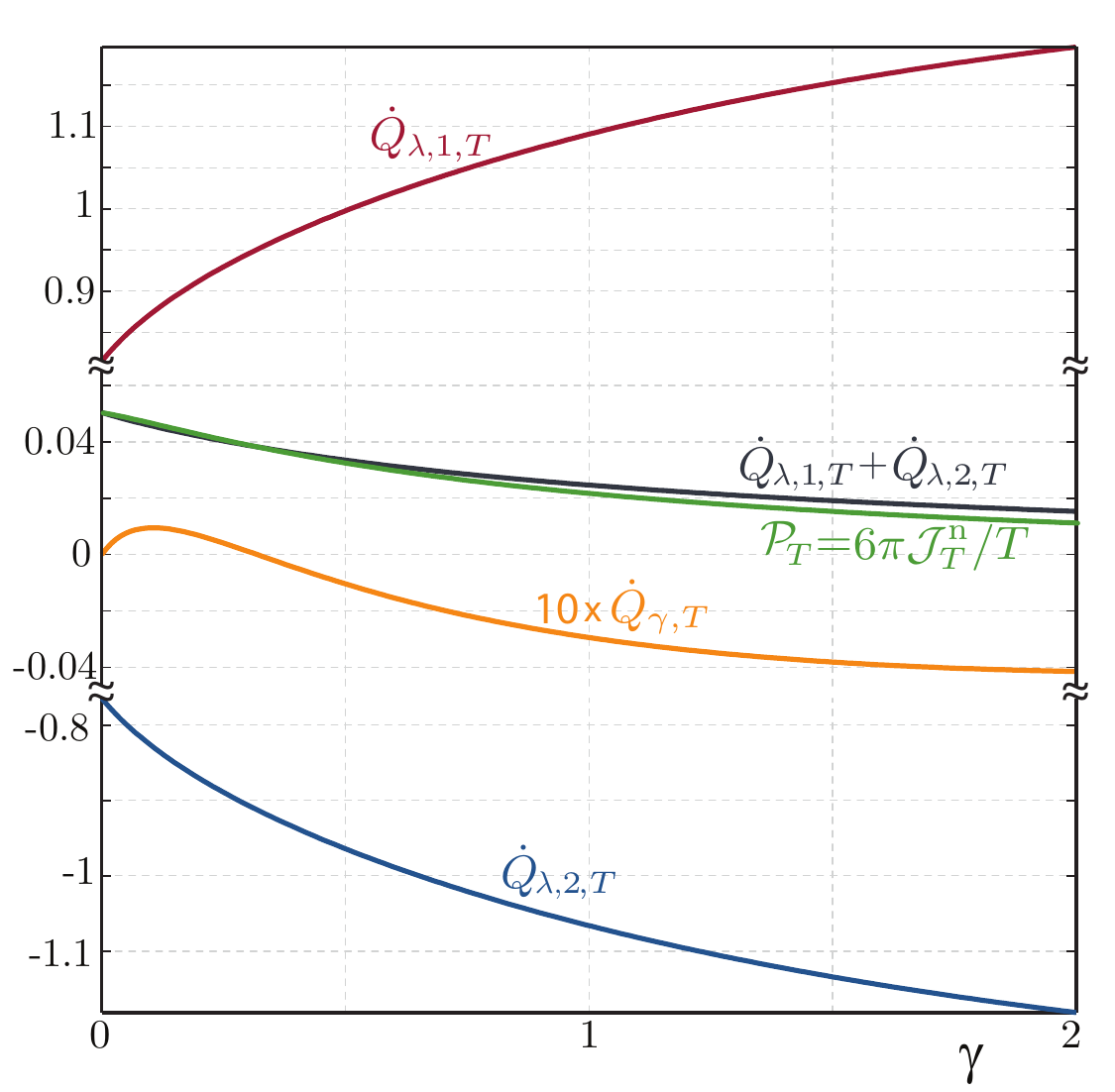}
\caption{(a) Particle current $\curpp$ and (b) spin current $\cursp$ vs. spin coupling $J_{xy}$ (dot-dashed line, $\gamma=0$; continuous line, $\gamma=0.5$; and dashed line, $\gamma=1$). (c) Average energy exchange rates $\dot{Q}_{\lambda,l,T}$, $\dot{Q}_{\lambda,1,T}+\dot{Q}_{\lambda,2,T}$, $\Pow_T$, and $\dot{Q}_{\gamma,T}$ vs. dephasing rate $\gamma$ for $J_{xy}=0.3$. Common parameters:  $\kappa=-2.5$, $h_z=15$, $\Lambda=0.2$, $\Delta p=0.5$, $\overline{p}=0.25$, and period $T=2$.
}
\label{FIG:fig4}    
\end{figure}

\subsection{Necessity of spin-coupling and spin-imbalance} \label{sec:necessity}

Figure~\ref{FIG:fig2} and \reff{fig3} indicate that the particle current (and hence the power) vanishes when either $J_{xy}=0$ (for any value of $\Delta p$) or $\Delta p=0$ (for any value of $J_{xy}$). In these cases we can show that the periodic steady state $\rhops$ is time-independent and given by the tensor product of density operators living only in the space of each spin or of the particle 
\begin{align}
\hat \rho_0=\hat \rho_0^{\mbox{\tiny spin,1}} \otimes \hat \rho_0^{\mbox{\tiny spin,2}} \otimes  \rhop_0^{\mbox{\tiny particle}}\label{eq:specsolu}. 
\end{align} 
Indeed, the reduced density operator for the particle is given by $\rhop_0^{\mbox{\tiny particle}}=\left[\hat{\mathbbm 1}/3 \right]$, for which $\Dop_{\gamma}(\rhop_0^{\mbox{\tiny particle}})=0$. 
Moreover, the individual reduced spin-density operators $\hat \rho_0^{\mbox{\tiny spin,l}}$ are of the explicit form given in \refe{spinDM_decoupled} and hence each of them is a steady state of the dissipators $\Dop_{\lambda,l}$, i.e., $\Dop_{\lambda,l}(\rho_0^{\mbox{\tiny spin,l}})=0$. 
The Hamiltonian contributions to the master equation can be decomposed into three parts, each of which vanishes since $[\Hops,\rhop_0]=0$, $[\Hopp,\rhop_0]=0$, and $[\Hopi,\rhop_0]=0$. Hence, $\Lind(\hat \rho_0,t)=0$ and thus $\hat \rho_0$ is the steady state for the $J_{xy}=0$ or $\Delta p=0$ cases. 

Since each spin is in the respective steady state of the local dissipator $\Dop_{\lambda,l}$, there are no energy exchanges with the baths, $\dot{Q}_{\lambda,l}(t)=0$, and also $\dot{Q}_{\gamma}=0$. Therefore, for $J_{xy}=0$ or $\Delta p=0$, the instantaneous power vanishes, $\Pow=0$ .    

\subsection{Non-separability of the periodic steady state}

For the general case of non-zero $J_{xy}$ and $\Delta p$, the periodic steady state is no longer of a separable product form. To demonstrate and quantify this, we perform a singular value decomposition of the periodic steady state between the spins subsystem and the particle subsystem at instant $t$, leading to a form $\rhops(t)=\sum_j s_j(t)   \; \hat B_j^{\mbox{\tiny spins}}(t) \otimes \hat B_j^{\mbox{\tiny particle}}(t)$. Here, $s_j(t)$ are the unique, positive singular values and $\hat B_j^{\mbox{\tiny spins/particle}}(t)$ are operators living on the spin and particle space, respectively, each of which is normalized to unity with respect to the Hilbert-Schmidt norm. 

\begin{figure}
\includegraphics[width=\columnwidth]{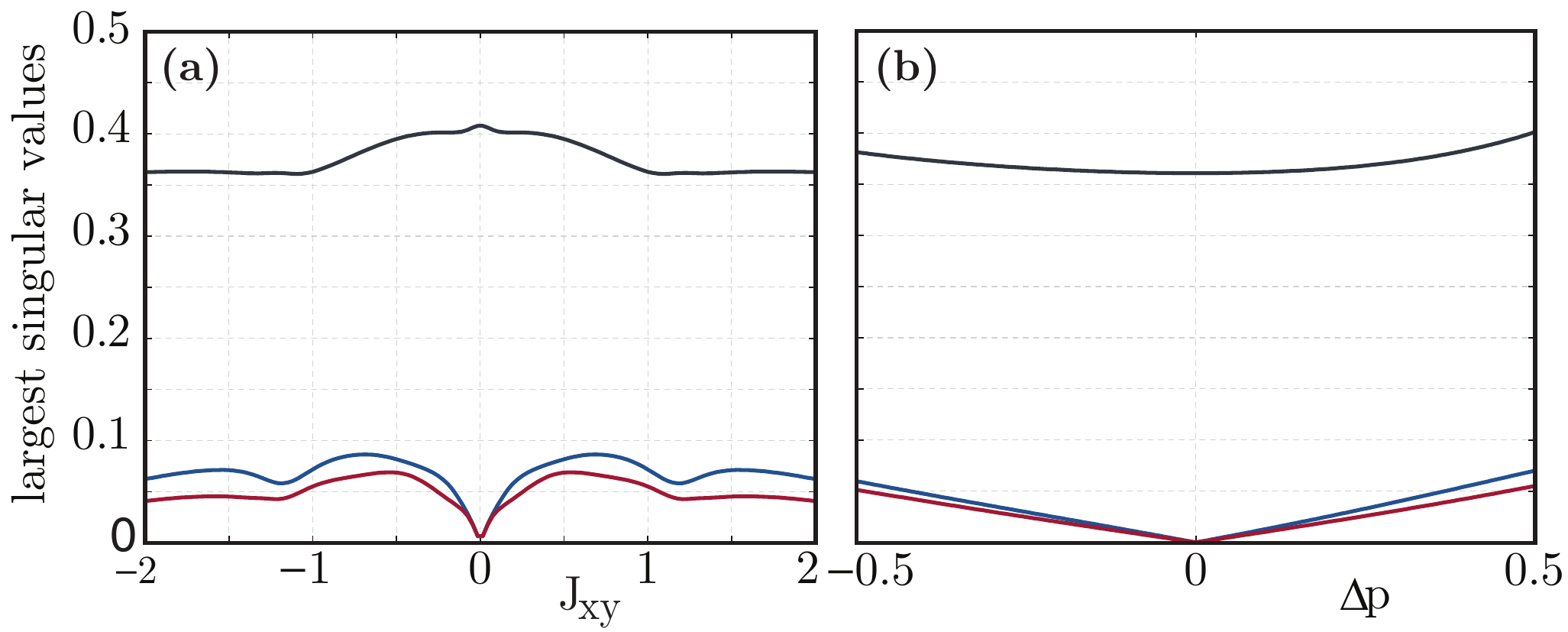}
\caption{Time-averaged singular values from the decomposition of the periodic steady state's singular value decomposition between the particle and spins subsystem vs. $J_{xy}$ (a) and $\Delta p$ (b). Common parameters: $\kappa=-2.5$, $h_z=15$, $\gamma=0$, $\Lambda=0.2$, $\overline p=0.25$, $T=2$ while $\Delta p=0.5$ in (a) and $J_{xy}=0.3$ in (b).}
\label{FIG:fig5}    
\end{figure}   

We refer to $\rhops(t)$ as separable iff, for all times, there is only a single non-zero singular value, i.e., $s_j(t)=\delta_{j,1} s_1(t)$, where $\delta_{j,1}$ is the Kronecker $\delta$. 
In \reff{fig5} we show the three largest time-averaged singular values $s_{j,T}=\frac 1 T \int_0^T s_j(t) \, dt$ as functions of $J_{xy}$ and $\Delta p$. Note that since the density matrix is normalized such that $\tr\!(\rhop) =1$ (and not with respect to the Hilbert-Schmidt norm corresponding to the standard Euclidean vector norm), the sum $\sum_j \left[s_j(t)\right]^2$ may generally differ from $1$. We generally observe that only for $J_{xy}=0$ or $\Delta p=0$ is the density matrix is exactly separable between spins and particle. In this decomposition, a vanishing of the time-averaged $s_{j,T}=0$ implies that $s_j(t)=0$ for all times, since the instantaneous singular values $s_j(t)$ are strictly non-negative. Hence, in general, the periodic steady state cannot be treated as separable of the spins and the particle, analogous to the case discussed in Ref.~\cite{TeoPoletti2016}.

\section{Conclusions} \label{sec:conclusions}  

We have proposed a minimal motor to perform work against a periodic driving from an out of equilibrium energy flow. Coupled-spin magnetization and particle transport is the key ingredient of this motor which is robust against a dissipative particle dephasing. 
On an analytic level, we have shown that coupling between the spins is necessary to achieve particle transport. However, stronger spin coupling prevents an effective conversion into powering the motion of the particle. 

We point out that using local Lindblad baths, it is not possible to associate a Boltzmann temperature from the relation $e^{-2h_{z,l}/k_B T_l}=p_l/(1-p_l)$, even if, as in our case, the Zeeman energy splitting $2 h_{z,l}$ is the dominant energy scale of the system. In fact, as shown in Ref.~\cite{LevyKosloff2014}, the use of local Lindblad baths may result in apparent violations of the second law of thermodynamics. 
It is thus important to stress that we are dealing with non-equilibrium, magnetization baths as described in Refs.~\cite{MeierLoss2003, BenentiRossini2009, SchuabLandi2016}. Moreover, in regimes in which particle current and energy exchanged with the baths are very small, more accurate master equations may be required, as discussed, for example, in Refs.~\cite{SaitoMiyashita2000, WichterichMichel2007, ThingnaHanggi2013, XuWang2016, GrifoniHanggi1998}.  

This system could be implemented in various set-ups with effective spin-$1/2$ systems made with ultracold ions in microtrap arrays \cite{Bermudez2013,Bermudez2011,Porras2004}, as well as in solid state systems. In particular, for this latter case, it could be possible to engineer this set-up using five quantum dots. Two quantum dots would take on the role of the spins coupled to a bath. Each of these would also be coupled to one of the remaining three (or more) quantum dots which form the circuit \cite{cap, Rogge2008, Thalineau2012, Seo2013}.

Future work could focus on analyzing the effect of system size, interactions, particle statistics and the role of measurement on the motor's performance. Our minimal model can be readily extended to investigate such effects.\\

{\it Acknowledgments:} We are grateful to J.M. Arrazola, F. Giazotto, J. Gong, M. Governale, A. Roulet and F. Taddei for fruitful discussions. This material is based on work supported by the Air Force Office of Scientific Research under Award No. FA2386-16-1-4041. D.P. acknowledges funding from the Singapore MOE Academic Research Fund Tier-2 project (Project No. MOE2014-T2-2-119, with WBS No. R-144-000-350-112), together with U.B., and from SUTD-MIT IDC (Project No. IDG21500104), together with C.T.

\begin{appendix}

\end{appendix}



\begin{thebibliography}{10}
\bibitem{sinedriving} Similar results can be obtained with a sinusoidal driving of the type $\phi(t)=\phi_0 + \phi_1 \sin(\omega t)$.
\bibitem{totalrho} Note that we use the full $\rhops(t)$, of both spins and particle, and not a reduced part of it.
\bibitem{direction} If the direction of energy flow between the system and bath reverses within one period, then this definition of absorbed heat differs from the net heat absorbed from one bath (and leads to a lower efficiency). However, for the parameters studied we have not observed such change of heat current direction.
\end{thebibliography}
\end{document}